\documentclass[12pt,epsf]{article}
\usepackage{graphicx, amsmath}

\begin{document}

\sloppy
\begin{flushright}{SIT-HEP/TM-46}
\end{flushright}
\vskip 1.5 truecm
\centerline{\large{\bf Curvatons and inhomogeneous scenarios with 
deviation from slow-roll}}
\vskip .75 truecm
\centerline{\bf Tomohiro Matsuda\footnote{matsuda@sit.ac.jp}}
\vskip .4 truecm
\centerline {\it Laboratory of Physics, Saitama Institute of Technology,}
\centerline {\it Fusaiji, Okabe-machi, Saitama 369-0293, 
Japan}
\vskip 1. truecm
\makeatletter
\@addtoreset{equation}{section}
\def\theequation{\thesection.\arabic{equation}}
\makeatother
\vskip 1. truecm

\begin{abstract}
\hspace*{\parindent}
The spectral index is studied at the point where scalar fields
 deviate from slow-roll during inflation. 
Considering the deviation that may cause a significant difference to
the time derivative of the Hubble parameter and also to the terms in
the evolution equation,
we show how the deviation
 affects the spectral index of the curvature perturbations. 
Considering conventional inflation, curvatons and other inhomogeneous 
scenarios as mechanisms for generating the cosmological perturbation,
 we examine whether the spectral index induced by the deviation from the
 standard slow-roll can explain the spectral index  $n-1>0$ at
 $k=0.002$/Mpc while keeping $n-1<0$ at a smaller scale.
\end{abstract}

\newpage
\section{Introduction}

\begin{figure}[h]
 \begin{center}
\begin{picture}(400,230)(0,0)
\resizebox{14cm}{!}{\includegraphics{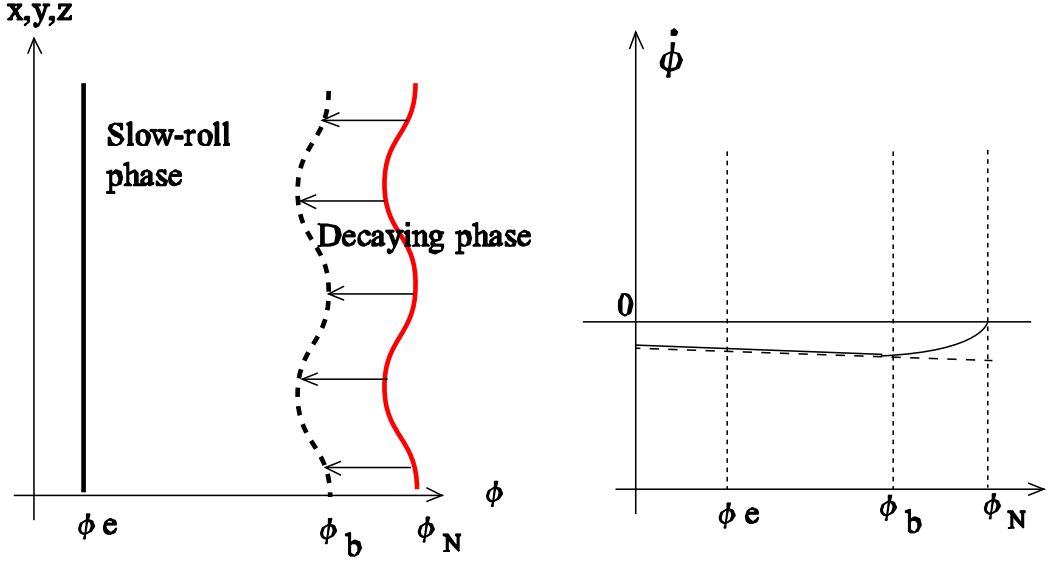}} 
\end{picture}
\label{fig:basic}
 \caption{In the traditional inflationary scenario, $\delta N \simeq H\delta (\Delta t)$ is caused 
 by  the fluctuation of the distance between the start-line at $\phi_N$
 (horizon crossing) and the goal-line at $\phi_e$ (end of inflation).
In the right-hand image, we show the inflaton velocity (continuous line)
with a small deviation from the slow-roll 
  velocity (dotted line) during the decaying phase between $\phi=\phi_N$
  and $\phi=\phi_b$.}
 \end{center}
\end{figure}

While the traditional inflationary scenario based on the slow-roll
approximation is considered to be broadly accurate \cite{EU-book},
there may be some shift from the slow-roll
trajectory \cite{deviation_from_slow} during inflation.
Besides the possibility of oscillating inflation \cite{Osc-infaltion},
deviation from the slow-roll may occur in the earliest stage of
inflation or just after the inflaton experiences a gap in the
single-field or hybrid-type potential.\footnote{
The same deviation may appear for other light scalar fields such as
curvatons and moduli fields. In Sec. 3, we consider similar situations
for these alternatives.} 
Allowing a short period of deviation from the slow-roll, the most
significant effect may occur when the inflaton stops during inflation,
where the inflaton turns around. Expressing the curvature perturbation
for slow-roll inflation as
\begin{equation}
\label{first1}
{\cal R}_k\simeq-\frac{H}{\dot{\phi}}\delta \phi_k,
\end{equation}
the divergence at the turnaround is brought about by $\dot{\phi}$ in the
denominator. 
This issue has been studied by Seto et. al. in Ref. \cite{SYK-stop},
using the standard formalism of cosmological perturbations.
They found that Eq. (\ref{first1}) is not correct when inflation stops,
but that the correct answer can be obtained by replacing $\dot{\phi}$
 by the slow-roll velocity $\dot{\phi}_s\equiv -V_{\phi}/3H$, 
where $V_\phi$ is the derivative of the potential with respect to the
inflaton.
The correct answer is thus given by 
\begin{equation}
{\cal R}_k\simeq-\frac{H}{\dot{\phi}_s}\delta \phi_k.
\end{equation}
The form of the curvature perturbation is the same as that for
conventional slow-roll inflation.
The $\delta N$ formalism is very useful for understanding this result.
The number of e-foldings $N$ is given by
\begin{equation}
N(t_e)\equiv \int^{t_e}_{t_N} Hdt=\int_{\phi_N}^{\phi_e}\frac{H}{\dot{\phi}}d\phi,
\end{equation}
where $t=t_N$ is the time when the perturbation crosses the horizon.
Here, we have introduced $\phi_N\equiv \phi(t_N)$ and $\phi_e\equiv
\phi(t_e)$ for simplicity. 
Following the model considered in Ref. \cite{SYK-stop}, we split the
inflaton velocity as 
\begin{equation}
\dot{\phi} = \dot{\phi}_s+\dot{\phi}_d,
\end{equation}
where $\dot{\phi}_s$ satisfies the slow-roll condition while
$\dot{\phi}_d$ is the decaying velocity that follows
$\dot{\phi}_d\propto e^{-3H t}$.
Note that the time elapsed during the decaying phase (where
$\dot{\phi}_d\ne 0$ is significant) is determined
by the Hubble parameter $H$ and $\dot{\phi}_d$ at the horizon
crossing.\footnote{For simplicity, we 
do not consider the fluctuation of the inflaton velocity $\delta
\dot{\phi}$. 
See Ref. \cite{modulated-inflation} for more details.}.
In terms of the $\delta N$ formalism \cite{delta-N-basic},
 the time passed after the horizon crossing is given by
\begin{equation}
\Delta t \equiv \Delta t_{dec}+\Delta t_{slow} \equiv 
(t_b-t_N)+(t_e-t_b),
\end{equation}
where $\Delta t_{dec}$ and $\Delta t_{slow}$ are the time passed during
the decaying phase and the slow-roll phase, which are shown in
Fig.\ref{fig:basic}.
$\dot{\phi}_s$ is constant during the decaying
phase, consistent with the usual assumption in slow-roll inflation. 
If the evolution equation of the inflaton field is homogeneous during
the decaying phase, the fluctuation at the boundary $\delta
\phi_b$ appears as a parallel 
 displacement of $\delta \phi_N$.\footnote{Of course, $\dot{\phi}_d$
does not cross $\dot{\phi}_d=0$, but we can define the ``boundary''
at, for example, $|\dot{\phi}_d/\dot{\phi}_s|=0.1$. In this case, the time
elapsed between the turnaround and the boundary is calculated from 
$e^{-3H(\Delta t_{dec})}=0.1$. It is then easy to calculate
the width of the decaying phase.
Considering the evolution equation for the inflaton field,
the required conditions for the homogeneous evolution are:
(1) homogeneous initial velocity, (2) homogeneous $H$ 
and (3) homogeneous $V$.
From the assumption that the inflaton
velocity is homogeneous at $\phi=\phi_N$, we find homogeneous kinetic
energy for the inflaton. 
However, there may be inhomogenities in $V$,
which may be caused by the perturbation $\delta \phi_N$.
Therefore, our argument is true for a sufficiently flat potential, while
one must consider inhomogeneous evolution caused by $\delta V$ and
$\delta V_\phi$ if the inhomogenities in the evolution equation are
significant. 
In this paper, we consider the former case following
Ref. \cite{SYK-stop}. Note that our argument is not true for the model
in which the inhomogenities created by $\delta \phi_N$ are not negligible
in the evolution equation for the inflaton field. }   
As a result, the fluctuation of the
 time elapsed during inflation is given by 
$\delta (\Delta t) = \delta (\Delta t_{slow}+\Delta t_{dec})
 \simeq \delta (\Delta t_{slow})$,\footnote{
During the decaying phase, we find 
$\delta (\Delta N_{dec})=\delta(H \Delta t_{dec})=
\delta H (\Delta t_{dec})+
H \delta(\Delta t_{dec})$. The inhomogenities related to $\delta H$
is $\delta H (\Delta t_{dec})\simeq \left[\frac{1}{6H
M_p^2}\frac{dV}{d\phi}\delta \phi  
\right]\times \left[H^{-1}\right] \simeq \sqrt{\epsilon_\phi}H/M_p$,
which can be neglected.} 
which suggests that the decaying phase does not
 play significant role in generating $\delta N$. 
Therefore, $\delta N$ in this scenario must be
 calculated in the slow-roll phase, using the conventional slow-roll
 parameter of the potential at $\phi=\phi_b$. 
Here, the net fluctuation of the time $\delta (\Delta t)$
is not generated during the decaying phase.
This result indicates that the curvature perturbation is determined by
 the slow-roll parameter of the potential 
\begin{equation}
\epsilon_\phi\equiv \frac{M_p^2}{2}\left(\frac{V_\phi}{V}\right)^2,
\end{equation}
where $M_p$ is the reduced Planck mass.
Therefore, even if the inflaton stops ($\dot{\phi}=0$) when it 
turns around at $\phi=\phi_N$, the curvature perturbation does not
diverge since $\delta N$
does not depend on the inflaton velocity in the decaying phase.
The result obtained by using the $\delta N$ formalism is consistent with
that obtained from the usual perturbation theory \cite{SYK-stop}.

Although the final form of the curvature perturbation may look the same as
 conventional slow-roll inflation, we should remember that the Hubble
parameter at the horizon crossing depends on the inflaton velocity.
For a typical example, the difference is significant when we consider
the time derivative of the Hubble parameter ($\dot{H} 
\propto \dot{\phi}^2$), which depends explicitly on the deviation of the
inflaton velocity.
Since the amplitude of the field fluctuation is determined by the Hubble
parameter at the horizon crossing, the deviation from the slow-roll
may lead to significant $\dot{\phi}_d$-dependence of the spectral index,
since the spectral index measures the scale-dependence of the
perturbation.  
This would also be true for many other scenarios 
such as curvatons \cite{curvatons-paper, matsuda_curvaton}, inhomogeneous
(p)reheating \cite{IP-paper,alt-inhore, ip-trapinflation} or the hybrid
of the 
two scenarios \cite{hybrid-matsuda}, as far as the amplitude 
of the seed fluctuations of light fields is determined by the Hubble
parameter at the horizon crossing.
Based on this simple idea, we calculate the spectral index for the
cosmological perturbations of various models, especially when there is
a significant deviation from the slow-roll velocity.

\subsection{Inflaton at a turnaround}
In the standard inflation model, following the above argument, 
the primordial curvature perturbation ${\cal R}_k$ is equal
to  
$-H \delta \phi_N/\dot{\phi}_s$ evaluated at $\phi=\phi_b$.
This gives the spectrum of the curvature perturbation \cite{EU-book} as
\begin{equation}
{\cal P_R}=\left(\frac{H}{\dot{\phi}_s}\right)^2
\left(\frac{H}{2\pi}\right)^2=\frac{H^2}{8\pi^2M_p^2\epsilon_\phi}
\end{equation}
where the slow-roll velocity is $\dot{\phi}_s\equiv-V_\phi/3H$.
There is no significant difference in the form of the curvature
perturbation,
but the spectral index of the curvature perturbation given by
\begin{equation}
n-1 \equiv \frac{d \ln {\cal P_R}}{d\ln k}=
-\frac{d\ln \epsilon_\phi}{d\ln k} +2\frac{d \ln H}{Hdt}=-4\epsilon_\phi
+2\eta_\phi-2\epsilon_H,
\end{equation}
where the definitions of the slow-roll parameters are 
$\eta_\phi \equiv M_p^2V_{\phi\phi}/V$ and 
$\epsilon_H\equiv -\dot{H}/H^2=\dot{\phi}^2/2M_p^2H^2$, should be
different. 
Paying attention to the derivative\footnote{
Here we use the relation $k=aH$, which leads to
 $d\ln k =\frac{da}{a}+\frac{dH}{H}
=\left(H+\frac{\dot{H}}{H}\right)dt \simeq H\left(1-\epsilon_\phi
\frac{\dot{\phi}^2}{\dot{\phi}_s^2}\right)dt\simeq H dt$.}
\begin{equation}
\frac{d}{d\ln k}\simeq\frac{d}{H dt}=\dot{\phi} \frac{ d}{H d\phi},
\end{equation}
the spectral index at a turnaround is 
\begin{equation}
n-1 \simeq 0.
\end{equation}
This result suggests that $\dot{\phi}=0$ during inflation may not
explain $n-1>0$ at $k=0.002$/Mpc and
$n-1<0$ at $k=0.05$/Mpc \cite{WMAP-data}.
For example, the value of the spectral index \cite{WMAP-data} is
\begin{equation}
\label{run-in-2}
n_{0.002} = 1.21^{+0.13}_{-0.16}
\end{equation}
at $k=0.002$/Mpc and 
\begin{equation}
\label{run-in-5}
n_{0.05} = 0.948^{+0.014}_{-0.018},
\end{equation}
which cannot be satisfied even if the inflaton stops during inflation.

Before the inflaton stops during inflation, the velocity of the inflaton
may be much larger than the slow-roll velocity.
Therefore, we will consider a more general situation with the inflaton
velocity 
$\dot{\phi}\equiv - f_d \dot{\phi}_s$, $1/\sqrt{\epsilon_\phi}\gg 1\ge f_d >0$.
Paying attention to the derivative 
\begin{equation}
\frac{d}{d\ln k}=\frac{d}{H(1-\epsilon_\phi f_d^2)d t}
\simeq-f_d \dot{\phi}_s \frac{ d}{H d\phi},
\end{equation}
the equation for the spectral index
\begin{equation}
n-1 \equiv \frac{d \ln {\cal P_R}}{d\ln k}=
-\frac{d\ln \epsilon_\phi}{d\ln k} +2\frac{d \ln H}{Hdt}
\end{equation}
leads to 
\begin{equation}
n-1 \simeq -f_d\left(2\eta_\phi -4 \epsilon_\phi\right) -2 f_d^2 \epsilon_\phi,
\end{equation}
which may explain $n-1>0$ at $k=0.002$/Mpc for $f_d \le 1$.

\section{Curvatons and other alternatives}
\subsection{When inflaton deviates from slow-roll during inflation}
\label{2-1}

Contrary to the result obtained for standard inflation, 
we find that in the curvaton model, a positive $n-1$ at
$k=0.002$/Mpc is a natural consequence of $\dot{\phi}=0$.
The spectral index for the curvaton model is
given by \cite{curvaton-index}
\begin{equation}
\label{curvaton-index}
n-1 \simeq 2\eta_\sigma-2\epsilon_H.
\end{equation}
Here, $\epsilon_\sigma$ is disregarded since it is
typically very small compared to $\eta_\sigma$.\footnote{For the
curvaton velocity $\dot{\sigma}>\dot{\phi}$, we find 
$d\ln k \simeq H\left(1-\epsilon_\sigma
\frac{\dot{\sigma}^2}{\dot{\sigma}_s^2}\right)dt\simeq H dt$.
The last approximation is true if 
$\dot{\sigma}\ll \dot{\sigma}_s/\sqrt{\epsilon_\sigma}$,
which is consistent with the standard curvaton scenario that suggests
very small $\epsilon_\sigma$.}
We assume $\epsilon_\sigma$ and $\eta_\sigma$ are
constants during inflation (i.e., the deviation occurs only for the
inflaton field.).
Then, the spectral index of the curvature perturbation that is related
to the perturbation of the curvaton when the inflaton stops
($\dot{\phi}=0$) is 
\begin{equation}
n-1 \simeq 2\eta_\sigma,
\end{equation}
which may be positive at $k=0.002$/Mpc,\footnote{It is possible to
construct 
a curvaton model in which $\eta_\sigma$ is negative during
inflation \cite{hilltop-matsuda}. We will not consider the possibility in
this paper.} while the spectral index at a smaller scale is
\begin{equation}
n-1 \simeq 2\eta_\sigma  -2\epsilon_\phi,
\end{equation}
which must be negative.
Note that in this model the flip in the sign of $n-1$ 
is not caused by the gap in the  
slow-roll parameters of the potential.
The flip occurs because $\epsilon_H\propto \dot{\phi}^2$
disappears at the turnaround where the inflaton stops. 
For curvatons, the running spectral index in
(\ref{run-in-2}) and (\ref{run-in-5}) leads to the condition
$\epsilon_\phi>0.04$, which is not constrained by the CMB normalization.
Note that large $\epsilon_\phi$ is favorable for the curvaton scenario,
since a larger $\epsilon_\phi$ decreases the inflaton-induced curvature
perturbation.
Therefore, the scenario of a running spectral index is more natural in
the curvaton scenario. 

Besides curvatons, as discussed above, inhomogeneous
preheating is a possibile mechanism for the cosmological perturbation.
The mechanism is consistent with the non-linear
parameter that measures the non-Gaussianity of the spectrum.
It also allows inflation with low-scale gravity \cite{low-inflation}. 
Moreover, the mechanism can be used to generate the initial density 
perturbation of the curvatons \cite{hybrid-matsuda}.
Although the mechanisms for generating the curvature perturbation in
inhomogeneous preheating scenarios are quite
different from that in the curvaton model, there is no significant
difference in the form of the spectral index \cite{IP-paper}.
The $\eta$-parameter of the light field is negative in a
trapping inflation model \cite{ip-trapinflation}, but in other scenarios
a positive $\eta$ can be taken.
The spectral index in inhomogeneous preheating is
given by Eq. (\ref{curvaton-index}), which naturally leads to $n-1>0$ 
when inflaton stops.

\subsection{When curvatons deviate from slow-roll during inflation}

As discussed in Section 1, deviation from the slow-roll may occur
in the earliest stage of inflation or just
after the field experiences a gap in the single-field or hybrid-type
potential. 
Of course, the same deviation may appear for scalar fields other than
the inflaton field. In this section, we will consider the spectral index
when the curvaton has an initial velocity in the
``wrong  direction'' (i.e., opposite
to the potential gradient) in the earliest stage of inflation and then
stops during inflation.  
For simplicity, we disregard deviations that may also appear for
the inflaton field (i.e., we consider $\dot{\phi}=\dot{\phi}_s$ for the
inflaton) and assume the deviation $\dot{\sigma}\ne \dot{\sigma}_s$ for
the curvaton. 
The evolution of the fluctuation of the curvaton field is given in terms
of the multi-field perturbation of the field equation \cite{multi-WBMR}:
\begin{equation}
\label{curv-eq}
\ddot{\delta \sigma}+3H\dot{\delta \sigma}+ 
\left[\frac{k^2}{a^2}+m_\sigma^2-\frac{1}{M_p^2 a^3}\frac{d}{dt}
\left(\frac{a^3\dot{\sigma}^2}{H}\right)\right]\delta \sigma
=\left[\frac{1}{M_p^2 a^3}\frac{d}{dt}
\left(\frac{a^3\dot{\phi}\dot{\sigma}}{H}\right)\right]\delta \phi,
\end{equation}
where we considered a standard quadratic potential
$V(\sigma)=\frac{1}{2}m_\sigma \sigma^2$ for the curvaton.
We find on large scales that
\begin{equation}
\label{multi-evo}
\ddot{\delta \sigma}+3H\dot{\delta \sigma}+ 
\left[m_\sigma^2-\frac{3\dot{\sigma}^2}{M_p^2}
+\frac{\dot{\sigma}^2}{M_p^2}\frac{\dot{H}}{H^2}
-\frac{2\ddot{\sigma}\dot{\sigma}}{M_p^2 H}\right]\delta \sigma
+\left[-\frac{3\dot{\phi}\dot{\sigma}}{M_p^2}
+\frac{\dot{\sigma}\dot{\phi}}{M_p^2}\frac{\dot{H}}{H^2}
-\frac{\ddot{\sigma}\dot{\phi}+\ddot{\phi}\dot{\sigma}}{M_p^2 H}
\right]\delta \phi=0.
\end{equation}
For an inflaton that satisfies $\dot{\phi}=\dot{\phi}_s$, 
the inflaton velocity $\dot{\phi}$ can be expressed in terms of the
slow-roll parameter $\epsilon_\phi$:
\begin{equation}
 \dot{\phi}=\dot{\phi}_s\equiv \sqrt{2\epsilon_\phi}HM_p.
\end{equation}
Let us first consider what happens when the curvaton stops during
inflation.
Considering $\dot{\sigma}=0$ and $\ddot{\sigma}=-V_\sigma$ and following
the 
standard assumption $\epsilon_\sigma \ll \eta_\sigma$ for the curvaton
potential, we find\footnote{Here we neglected the term related to
$\ddot{\delta \sigma}$ as in the usual calculation,
since for the terms in Eq. (\ref{multi-evo}),
 there is no significant change in the approximation even if the
 curvaton stops during inflation. 
We neglected the term related to 
$\ddot{\sigma}\dot{\phi} \delta \phi$ assuming that 
$\sqrt{\epsilon_\sigma\epsilon_\phi} \ll \eta_\sigma$. } 
\begin{equation}
H^{-1}\dot{\delta \sigma} \simeq -\eta_{\sigma}\delta\sigma.
\end{equation}
This equation gives the evolution of $\delta \sigma$ after the horizon
crossing. 
In addition to the scale-dependence of the curvaton fluctuation $\delta
\sigma$, which can be seen from the evolution equation, we must consider
the variation in the initial value.
Considering these effects, we find that the spectral index for the
curvaton when the curvaton stops ($\dot{\sigma}=0$) is given by
\begin{equation}
n-1=2\eta_\sigma-2\epsilon_H.
\end{equation}
In this case, the running is very small since the slow-roll parameter
$\epsilon_\sigma$ is typically very small.

Besides the possibility of $\dot{\sigma}=0$, the
running of the spectral index may be significant if the 
curvaton has a significant velocity 
$|\dot{\sigma}|\gg |\dot{\sigma}_s|$.
Considering the deviation from the slow-roll, we split 
the curvaton velocity as
\begin{equation}
\dot{\sigma} = \dot{\sigma}_s+\dot{\sigma}_d \equiv f_d \dot{\sigma}_s,
\end{equation}
where $\dot{\sigma}_s$ satisfies the slow-roll condition while
$\dot{\sigma}_d$ is the decaying velocity that follows
$\dot{\sigma}_d\propto e^{-3H t}$.
Substituting the curvaton velocity, we find from Eq. (\ref{curv-eq})
that\footnote{The change in the slow-roll approximation may be 
significant if the deviation from the slow-roll is significant.
However, as far as we are considering $|n-1|\ll 1$, we may use the usual
approximation.
We neglected the term related to 
$\ddot{\sigma}\dot{\phi} \delta \phi$, since the term 
is smaller than the term proportional to $f_d^2$.}
\begin{equation}
H^{-1}\dot{\delta \sigma} \simeq 
-\left(\eta_{\sigma} +2f_d^2\epsilon_\sigma\right)\delta\sigma,
\end{equation}
where we have disregarded higher order terms 
and considered 
$\ddot{\sigma}\simeq -3H \dot{\sigma}$.
The spectral index for the curvaton is thus given by
\begin{equation}
n-1=4 f_d^2 \epsilon_\sigma+2\eta_\sigma-2\epsilon_H.
\end{equation}
Here $\epsilon_H$ is given by $\epsilon_H\equiv
\sum_i\dot{\phi_i}^2/2M_p^2H^2$, with the sum applying for 
all fields that have non-zero velocity during inflation.
Note that the significant deviation
($f_d^2>\epsilon_\phi/\epsilon_\sigma$)  
leads to $n-1>0$.

Our conclusion in this section is that in the standard curvaton scenario 
$\dot{\sigma}=0$ does not lead to a significant running of the spectral
index, which explains $n-1>0$ at $k=0.002$/Mpc, while the deviation
$\dot{\sigma}^2 > (\epsilon_\phi/\epsilon_\sigma)\dot{\sigma}_s^2$
can lead to a spectral index of $n-1>0$ with significant running.
Large $\epsilon_\sigma$ is required in order to explain the running of
the spectral index with $\dot{\sigma}=0$, which suggests that 
the curvaton scenario is not standard in this case.

\section{Conclusions and discussions}
We have studied the spectral index at a scale when inflaton
deviates from slow-roll during inflation. 
It might be suspected that the curvature power spectrum would be
considerably enhanced when the inflaton stops during inflaton.
Following the discussions in Ref. \cite{SYK-stop}, the formula for the
spectrum of the density perturbation derived at $\dot{\phi}=0$ seems to 
contain a solution singular at $\dot{\phi}=0$.
However, the solution
is regular, as has been discussed in Ref. \cite{KH-reg}.
The enhancement may appear in a decaying mode of the curvature
perturbation, but it does not appear in the nondecaying mode.
Therefore, there is a significant enhancement only for the
short-wavelength perturbations.
The important result in Ref. \cite{SYK-stop} is that the final form of
the curvature perturbation at large scales is the same as that
obtained for conventional slow-roll inflation, even if the inflaton stops
during inflation.
The $\delta N$ formalism is very useful for understanding why the
deviation 
from the slow-roll does not alter the final form of the curvature
perturbations at large scales.

Also of importance is the fact that although the final forms of the
curvature perturbation look quite similar, there is still a significant
difference from the standard inflation.
For example, note that $\dot{H}$ is proportional to $\dot{\phi}^2$,
which shows that the time-derivative of the Hubble parameter is
significantly different from that for standard inflation.
The differences lead to a crucial disparity in the spectral index.
As a result, a significant difference appears in the time derivative of
the seed-field perturbation, which causes a significant modification of
the spectral index.
Our results show that the modification is significant in the standard
inflation model.
Interestingly, the running of the spectral index is also significant in
other alternatives such as curvatons and inhomogeneous scenarios.

In addition to the deviation $\dot{\phi}\ne \dot{\phi}_s$, we considered
a similar situation for the curvaton field ($\dot{\sigma}\ne 
\dot{\sigma}_s$), which leads again to a significant running of the
spectral index. 

Despite the characteristic behavior of the scale-dependent spectrum that
is expected in these models, the actual resolution of the current CMB
data with respect to the scale dependence is not accurate enough to
determine the origin of the spectrum in terms of the scale-dependence.
 If future observations provide more accurate data with respect to the
 scale dependence, we hope the signature of the initial state of the
 Universe may be detected in the deviation from the slow-roll.

\section{Note added}
The model discussed in Sec.\ref{2-1} may be similar to DBI-curvatons
discussed in Ref.\cite{DBI-Curvaton}.
They considered two-field DBI action for the fields $\{\phi, \sigma\}$ and
find that the sound velocity can be very small when $\dot{\phi}\gg
\dot{\sigma}$ for the inflaton $\phi$, which leads to some remarkable
properties for the curvature perturbations caused by the curvaton 
field $\sigma$.

\section{Acknowledgment}
We wish to thank K.Shima for encouragement, and our colleagues at
Tokyo University for their kind hospitality.

\end{document}